\documentclass{PoS}
\usepackage{caption}
\usepackage{epsfig,amsmath}

\title{Extending parton branching TMDs to small $x$}

\ShortTitle{Extending PB TMDs to small $x$}

\author{\speaker{Sara Taheri Monfared}\\
        DESY, Hamburg, FRG\\
        E-mail: \email{sara.taheri.monfared@desy.de}}

\author{Francesco Hautmann\\
        Elementary Particle Physics, University of Antwerp, B 2020 Antwerp\\
        RAL, Chilton OX11 0QX and University of Oxford, Oxford OX1 3NP\\
        UPV/EHU, University of the Basque Country, E48080 Bilbao\\
        E-mail: \email{hautmann@thphys.ox.ac.uk}}

\author{Hannes Jung\\
        DESY, Hamburg, FRG\\
        E-mail: \email{hannes.jung@desy.de}}
        
\author{Melanie Schmitz\\
        DESY, Hamburg, FRG\\
        E-mail: \email{melanie.schmitz@desy.de}}

\abstract{We explore the possibility to include small-$x$ dynamics effects  in the parton branching (PB) approach  to 
transverse momentum dependent (TMD) parton distribution functions. To this end,   we   first    revisit the PB method at leading order, presenting a 
new fit to inclusive-DIS precision data, and performing a numerical study of the dynamic soft-gluon resolution scale. Next we investigate the effects of 
modified CCFM kernels, including both Sudakov and non-Sudakov form  factors.}

\FullConference{XXVII International Workshop on Deep-Inelastic Scattering and Related Subjects - DIS2019\\
		8-12 April, 2019\\
		Torino, Italy}

\begin{document}

\section{Introduction}
\label{sec1}

In Refs.~\cite{Hautmann:2017fcj,Hautmann:2017xtx}  a parton branching (PB) equation has been proposed 
for the evolution of transverse momentum dependent (TMD) parton distribution functions.  
In this   approach, soft gluon emission and transverse momentum recoils  in the QCD evolution  are treated 
by introducing the soft-gluon resolution scale $z_M$ to separate 
resolvable and non-resolvable branchings,   and the Sudakov form factors $\Delta_a$ (where  
$a = 1 , \dots , 2 N_f + 1 $ is  the flavor index,  with $N_f$  the number of quark  flavors) 
 to express  the probability for no resolvable  branching 
in a given evolution interval. 

The purpose of the study reported in this article is to extend the PB  TMD approach to include effects of small-$x$ dynamics. The study consists of two parts:  
a numerical  analysis  of the PB method at leading order (LO) in the strong coupling $\alpha_s$, for  which in particular  we carry out  a detailed  investigation of the 
role of the soft-gluon resolution scale $z_M$; an extension of the PB evolution kernels to the small-$x$ region  by including, 
besides the Sudakov form factors, also the non-Sudakov form 
factors according to  CCFM  methods~\cite{skewang,Catani:1989sg,hj-ang}.  

The article is organized as follows. In Sec.~2 we describe the LO PB analysis,  
performing a new fit to inclusive-DIS precision data~\cite{Abramowicz:2015mha}. In Sec.~3 we study the dependence of the fit on the parameter $q_0$, 
 which controls the minimum transverse momentum emitted in the branching evolution and is related to the soft-gluon resolution  scale $z_M$  according to 
 angular ordering~\cite{Hautmann:2017fcj,eswbook}. In Sec.~4 we introduce  modified CCFM kernels  and investigate their effect on the 
results of parton evolution. We give conclusions in Sec.~5. 

\section{PB method at LO}

The PB method has been developed in~\cite{Hautmann:2017fcj,Hautmann:2017xtx} evaluating the evolution kernels up to next-to-leading order (NLO). 
These  results have been used in Ref.~\cite{Martinez:2018jxt} to perform NLO fits  to precision HERA1+2 measurements~\cite{Abramowicz:2015mha}   
and in Ref.~\cite{Martinez:2019mwt}, along with NLO matrix elements for Drell-Yan  hadroproduction, to make predictions for $Z$-boson transverse 
momentum spectra  at the LHC.  
 
Our goal  in this work is  to explore the extension of the PB method to include small-$x$ CCFM effects and 
for this purpose it is useful to start from  the formulation of the PB method at LO.   To this end, we have performed an analysis at LO analogous to that 
of Ref.~\cite{Martinez:2018jxt} and obtained PB TMD fits at LO.    These LO fits may be useful in their own right for further applications.   

The fits    are performed 
using the  open-source fitting platform   \verb+xFitter+~\cite{Alekhin:2014irh} and the numerical techniques  
developed in~\cite{Hautmann:2014uua,hj-updfs-1312} to treat the transverse momentum dependence in the fitting procedure. 
Similarly to Ref.~\cite{Martinez:2018jxt}, 
we consider the two angular ordered TMD sets (Set1 and Set2) using   two different choices for the $\alpha_s$ evolution. 
 We consider the same functional forms for initial densities and proceed with the LO approximation for an extension towards the LO CCFM evolution.
The integrated TMD is fitted within  \verb+xFitter+   at LO, where the PB method is implemented, to precision HERA1+2 measurements in the range of $3.5< Q^2<50000$ GeV$^2$ and $4.10^{-5}<x<0.65$ \cite{Abramowicz:2015mha}. Our  $\chi^2/dof$ for both sets are reasonably good (1.24 and 1.26 for set1 and set2, respectively) and comparable to the NLO fit.

\section{Soft-gluon resolution effects}

Partons emitted with a transverse momentum smaller than a certain value given by the resolution scale cannot be resolved.
The resolution scale $z_M$ can change dynamically with the evolution scale. For angular ordering condition we consider 
$z_M=1-\frac{q_0}{q_i}$, while $q_0$ is chosen to be very small ($q_0=0.01$ GeV) to ensure that the PB evolution coincides with the standard DGLAP evolution.

\begin{figure}[h]
\centering \includegraphics[width=0.65\textwidth]{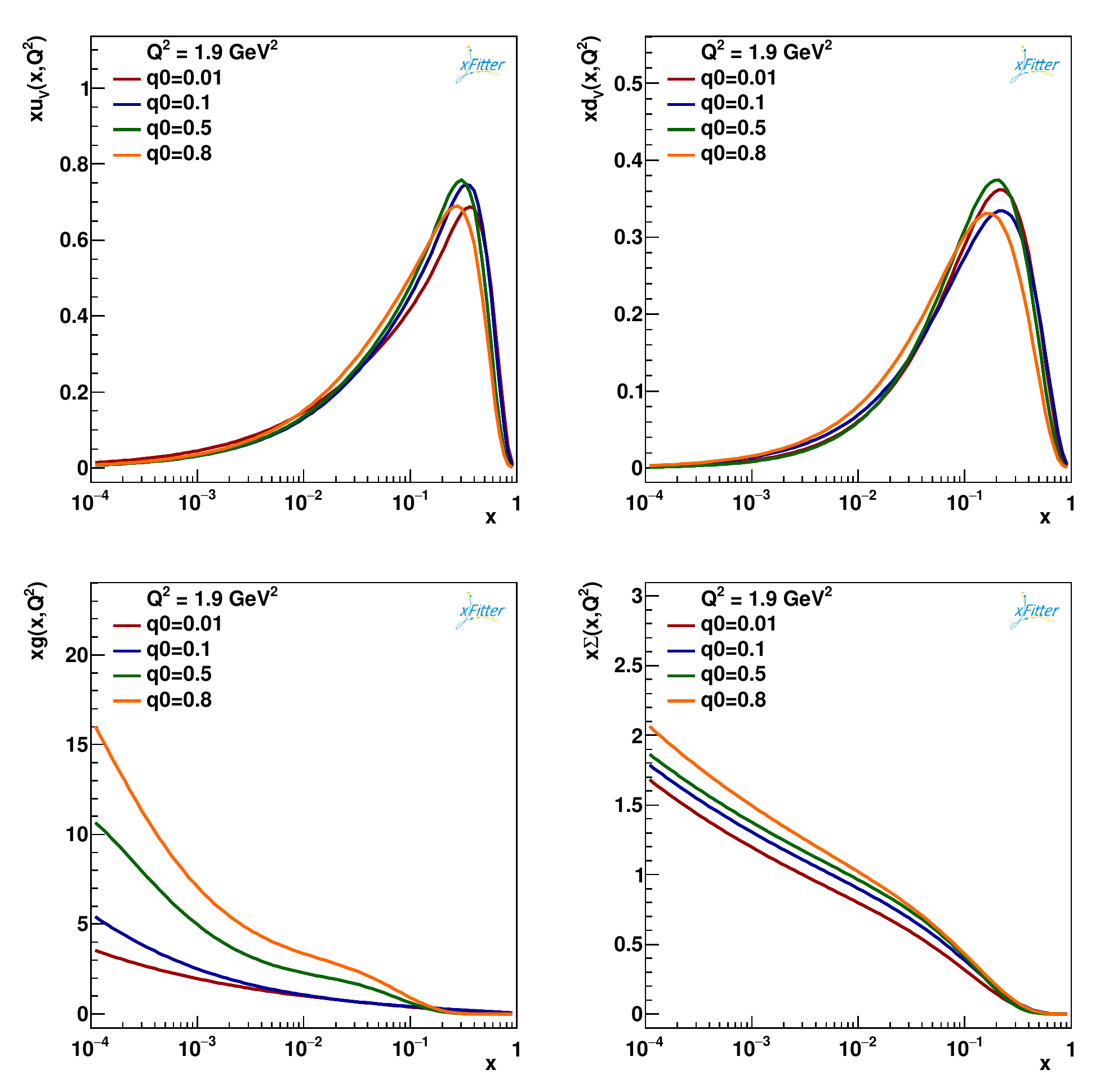}
\caption{Parton densities for different values of $q_0$ at $Q^2=1.9$ GeV$^2$}
\label{Fig:dynamiczm}
\end{figure}

We perform fits to DIS data with different choices of $q_0$. In Fig. \ref{Fig:dynamiczm} we illustrate how much would be the actual distribution changed by varying $z_{M}$. 
We observe that with higher $q_0$ (smaller $z_M$), the gluon distribution becomes significantly larger, essentially in the small $x$ region and the $\chi^2/dof$ also increases.

We investigate further how the choice of $Q_{min}^2$, the minimum $Q^2$ of the data within the fit, influences the quality of fit. We perform several fits with the different $Q_{min}^2$. As illustrated in Fig \ref{Fig:dynamiczm}, a reasonable fit can be determined for larger $q_0$ if $Q^2_{min}>10$ GeV$^2$. 
\begin{figure}
\centering \includegraphics[width=0.6\textwidth]{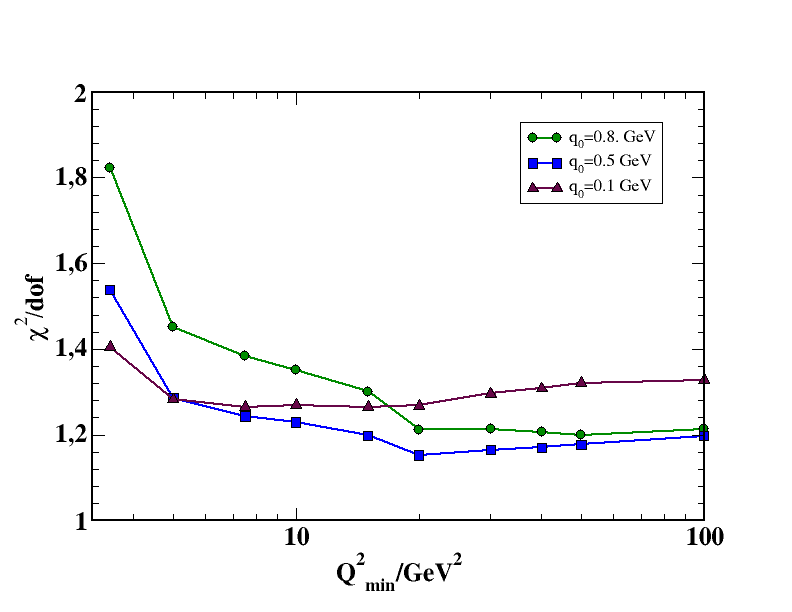}
\caption{The dependence of $\chi^2/dof$ on $Q_{min}^2$ of LO fit for $q_0=0.1,0.5,0.8$ GeV}
\label{Fig:dynamiczm2}
\end{figure}

\section{Extension to include small $x$ processes}
In this section we describe how the full  angular ordering condition enlarges the phase space. We go from the DGLAP ordering (${q}_{i}>q_{,i-1}$) to the 
angular ordering (${q}_{i}>z_{i-1} q_{i-1}$). 
Here $q_{i}$ are the rescaled transverse momenta.
In terms of the transverse momentum in the $t$-channel, DGLAP ordering means $k_\perp<{q}_{i}$, while angular ordering also covers $k_\perp>{q}_{i}$. 
This is due to the coherent effects.  
 We need to introduce a non-Sudakov form factor to sum all virtual corrections in which the rescaled transverse momenta of on-shell emitted gluons are smaller than the $k_\perp$ ($k_\perp>q_i$).
The modified CCFM  splitting functions at LO including the non-Sudakov form factor are given by 
 \begin{eqnarray}
&&P_{gg}^{(0)}=6\;(\frac{\alpha_s}{2\pi})\;(\frac{1}{z}{\widetilde{\Delta}_{ns}}+\frac{1}{1-z}-2+z(1-z)) \; \widetilde{\Delta}_s ~,\nonumber\\
&&P_{gq}^{(0)}=\frac{4}{3}\;(\frac{\alpha_s}{2\pi})\;(z-2+\frac{2}{z}{\widetilde{\Delta}_{ns}}) \; \widetilde{\Delta}_s~,\nonumber\\
&&P_{qg}^{(0)}=\frac{1}{2}\;(\frac{\alpha_s}{2\pi})\;(z^2+(1-z)^2) \; \widetilde{\Delta}_s~, \nonumber\\
&&P_{q_iq_i}^{(0)}=\frac{4}{3}\;(\frac{\alpha_s}{2\pi})\;(\frac{1+z^2}{1-z})\; \widetilde{\Delta}_s~.
\end{eqnarray}
with $\widetilde{\Delta}_s$ and ${\widetilde{\Delta}_{ns}}$ defined as
\begin{eqnarray}
&&\widetilde{\Delta}_s=exp\left (- \int_{z_{i-1}q_{i-1}} ^{{q_i}}\frac{d{q}'^{2}}{{q}'^{2}} \int^{z_M} dz\; \frac{1}{1-z}\right )~,\nonumber\\
&&{\widetilde{\Delta}_{ns}}=exp\left (- \int_{z_{i-1}q_{i-1}} ^{{k_\perp}}\frac{d{q}'^{2}}{{q}'^{2}} \int^{z_M} dz\; \frac{1}{z}\right )~.
\end{eqnarray}
The non-Sudakov form factor is relevant only when the emitted gluon is fast and therefore it is important only in presence of the $1/z$ term in the DGLAP splitting function.  
Therefore we included them in $P_{gg}$ and $P_{gq}$ splitting functions. 

In the next step, we re-write the CCFM splitting functions in terms of the full DGLAP splitting function with the DGLAP Sudakov form factor: 
\begin{equation}
\widetilde{\Delta}_s \rightarrow \Delta_s=exp\left (- \int_{z_{i-1}q_{i-1}} ^{{q_i}}\frac{d{q}'^{2}}{{q}'^{2}} \int^{z_M} dz\; (\frac{1}{z}+\frac{1}{1-z}-2+z(1-z))\right )~.
\end{equation}
We observe that $\frac{1}{z}\Delta_{ns}\Delta_s$ is already covered by $\frac{1}{z}\Delta_s$ for $k_\perp<{q}_{i}$ and the non-Sudakov form factor acts only if $k_\perp>q_i$. We obtain the following $\Delta_{ns}$ form factors:  
\begin{eqnarray}
&&\widetilde{\Delta}_{ns} \rightarrow \Delta_{ns}=exp\left (- \int_{q_{i}} ^{{k_\perp}}\frac{d{q}'^{2}}{{q}'^{2}} \int^{z_M} dz\; \frac{1}{z}\right )\;\;\;   \text{for}\; k_\perp>q_i\nonumber\\
&&\widetilde{\Delta}_{ns} \rightarrow \Delta_{ns}=1\;\;\;   \text{for}\; k_\perp<q_i
\end{eqnarray}
In Fig. \ref{Fig:,nonsud} the gluon density is shown as a function of $x$ for the evolution scale $\mu^2=100$ GeV$^2$ using the benchmark starting distribution with a black curve.
\begin{figure}[h]
\centering
\includegraphics[width=0.65\textwidth]{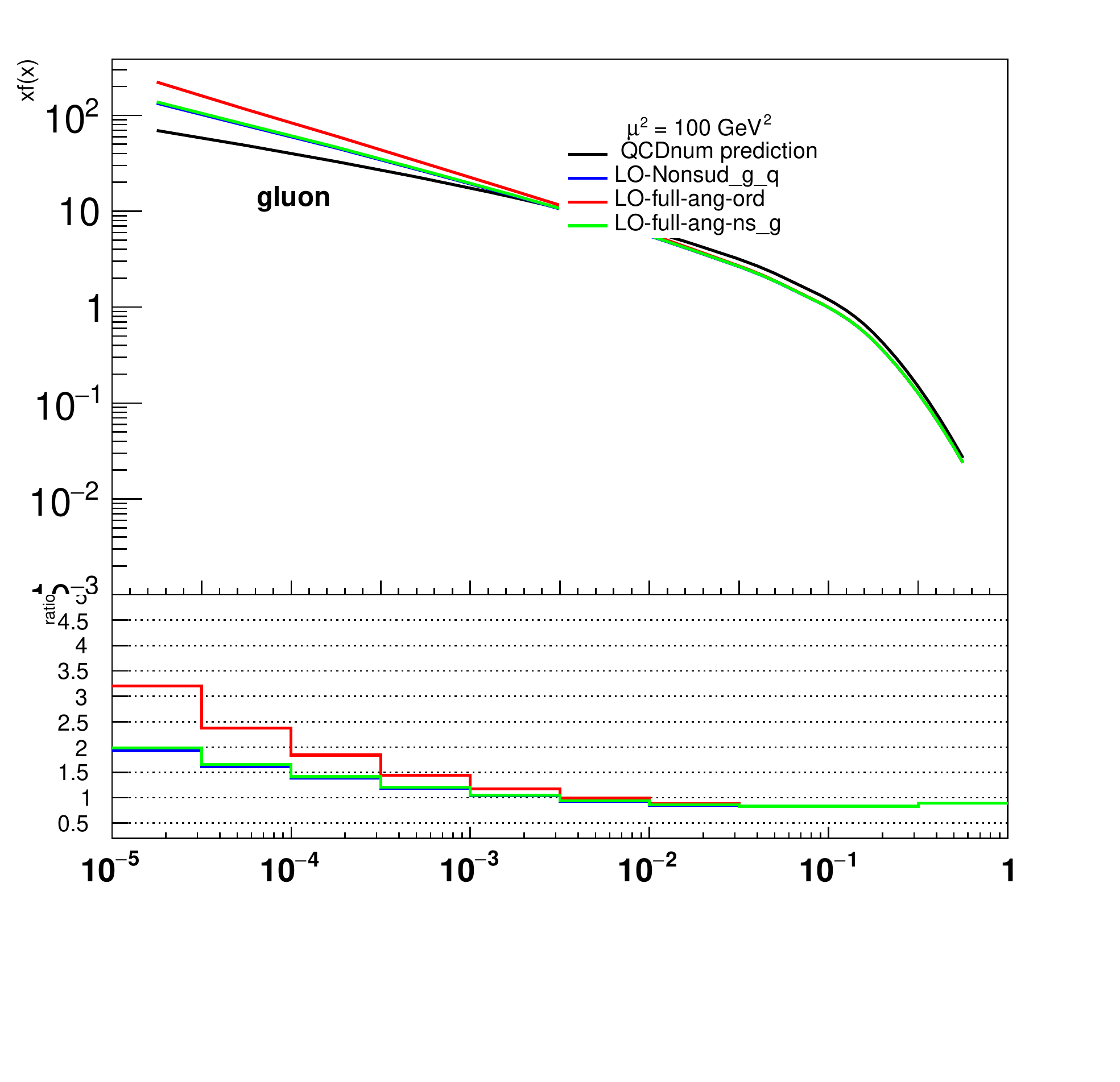}
\vspace{-1.5cm}
\caption{The black curve represents the QCDNUM prediction. The red curve includes full angular ordering. The non-Sudakov form factors are included in $p_{gg}$ and $p_{gq}$ in the green and blue curves, respectively.}
\label{Fig:,nonsud}
\end{figure}
 QCDNUM \cite{Botje:2010ay} agrees with PB method in the DGLAP limit as shown in ref. \cite{Hautmann:2017fcj}. 
The red curve shows the results when the phase space is opened up to include angular ordering. It allows for unordered emission and increases the gluon density at small $x$, essentially below $10^{-3}$.  
Including virtual corrections via non-Sudakov form factor (for  $k_\perp>q_i$) into $P_{gg}$ leads to small $x$ suppression (green curve in Fig.~\ref{Fig:,nonsud}). 
Including the non-Sudakov form factor also for $P_{gq}$ leads to the blue curve which is slightly below the green one. 
However, the gluon density at small $x$ is still larger compared to DGLAP based curve. 

This step by step study is possible via the PB method because it can solve the evolution equation via an iterative procedure.

\section{Conclusion}

TMD formalisms are relevant in QCD to describe contributions to the initial states of hadronic 
collisions both for  high $x$  and for   low $x$ (see e.g.~\cite{Angeles-Martinez:2015sea}).  
An essential ingredient is  provided by parameterizations and fits to experimental data 
for TMD distribution functions~\cite{tmdplott}. 

The PB method has been recently proposed  as a flexible approach which 
 can be applied to a wide range of processes and observables. 
 
 In this article we have performed new PB TMD fits at LO, and we have presented a method to 
 incorporate CCFM effects into the PB formulation. Our results  can be used to extend the existing 
 simulations of QCD cascades into the semi-hard regime.
\section{Acknowledgment}
STM thanks the Humboldt Foundation for the Georg Forster research fellowship. MS thanks DESY for giving the opportunity to go to the DIS conference.

\end{document}